\documentclass[aps,prb,twocolumn,superscriptaddress,showpacs]{revtex4}
\usepackage{graphics}
\usepackage{graphicx}
\usepackage[english]{babel}
\usepackage{color}
\usepackage{amsmath}

\begin{document}

\preprint{APS/123-QED}

\title{Second ac screening step as a probe for the first-order melting transition in layered
vortex matter at intermediate temperatures}

\author{Gonzalo Rumi}
\affiliation{Laboratorio de Bajas Temperaturas, Centro At\'{o}mico Bariloche, CNEA, Argentina.}
\affiliation{Instituto de Nanociencia y Nanotecnología, CONICET-CNEA, Nodo Bariloche, Argentina.}
\affiliation{Instituto Balseiro,
CNEA and Universidad Nacional de Cuyo, Bariloche, Argentina.}
\affiliation{Laboratorio Nacional de Haces de Neutrones, CNEA, Argentina.}

\author{Pablo Pedrazzini}
\affiliation{Laboratorio de Bajas Temperaturas, Centro At\'{o}mico Bariloche, CNEA, Argentina.}

\author{Hernán Pastoriza}
\affiliation{Sensores y dispositivos, Centro At\'{o}mico Bariloche, CNEA, Argentina.}
\affiliation{Instituto de Nanociencia y Nanotecnología, CONICET-CNEA, Nodo Bariloche, Argentina.}
\affiliation{Instituto Balseiro,
CNEA and Universidad Nacional de Cuyo, Bariloche, Argentina.}

\author{Marcin Konczykowski}%
\affiliation{Laboratoire des Solides Irradi\'es, CEA/DRF/IRAMIS, \'{E}cole Polytechnique, CNRS, Institut Polytechnique de Paris, F-91128 Palaiseau, France.}

\author{Yanina Fasano}

\affiliation{Laboratorio de Bajas Temperaturas, Centro At\'{o}mico Bariloche, CNEA, Argentina.}
\affiliation{Instituto de Nanociencia y Nanotecnología, CONICET-CNEA, Nodo Bariloche, Argentina.}
\affiliation{Instituto Balseiro,
CNEA and Universidad Nacional de Cuyo, Bariloche, Argentina.}

\date{\today}

\begin{abstract}
 We present a new probe for the first-order transition for layered vortex matter: A second step in the screening of an ac field that is independent of the frequency and amplitude of the excitation. This second step is observed in the intermediate temperature and field ranges where detecting the jump in induction associated with the transition is rather elusive with standard magnetometry techniques. We observe this second step following  a novel experimental protocol where the screening of a locally-generated ac field is remotely detected in another region of the sample. The coincidence of the typical temperature of the second  step in direct and remote measurements strongly supports this feature is a probe of the first-order transition. This nonlocal effect  detected at distances of thousands of vortex lattice spacings away indicates that a very efficient mechanism propagates the change in rigidity of the structure from the more (liquid) to the less (solid) symmetric vortex phases.


\end{abstract}

\maketitle


\section*{\label{sec:Intro}INTRODUCTION}

The first-order melting transition in systems with disorder remains a subject of active interest, as understanding its phenomenological details is important for both applications and basic research.~\cite{Chaikin,Kelton2010} 
 These transformations typically entail a global density jump,~\cite{Chaikin} and local probes in various condensed matter systems~\cite{Soibel2000,Anderson2002,Gasser2003,Duhan2025} reveal the complex nucleation and growth processes of the stable phases at a microscopic scale. First-order melting transitions also entail a loss of rigidity when the system transitions from a less symmetric phase to a more symmetric phase.~\cite{Chaikin} The efficiency with which one phase grows over another strongly depends on the  ability of the system to propagate rigidity from the less symmetric phase. One possible approach to explore this rigidity propagation is to remotely probe the response to temperature or density excitations generated elsewhere, at some lattice spacing distances. This type of nonlocal study can also provide valuable information on the location of the  transition line when regular global techniques cannot detect it.

The vortex matter in high-temperature superconductors provides a model system for studying this issue over a wide range of temperatures, densities, and disorder.~\cite{Blatter,LeDoussal} Moreover, considering extremely-layered superconductors also provides an opportunity to study the phenomenology of rigidity propagation in less symmetric phases for highly anisotropic, and thus softer, elastic systems.~\cite{Blatter} The first-order vortex melting transition involves a jump in magnetic flux density, $\Delta B$, detected both with bulk and local techniques.~\cite{FOT-Pastoriza,FOT-Zeldov,Safar1994,Sasagawa1998,Qiu1998,Soibel2000}  
This transition  separates a liquid phase with zero shear modulus\cite{Shear} at high temperatures from a solid phase at lower temperatures possessing the rigidity of a finite shear modulus and quasi-long-range positional order.~\cite{Cubitt1993,AragonSanchez2019} 
In extremely-layered  high-$T_{\rm c}$ superconductors, at the transition vortices decouple along the $c$-axis on warming.~\cite{Doyle1995,Righi1997,Matsuda1997,Goffman1998,Enriquez1998,Qiu1998,Ando1999,Blasius1999,Colson2003} 
The coupling of vortices on cooling produces a drag effect that recovers the rigidity of the system along the direction of vortices and gives rise to nonlocal effects.~\cite{Busch1992,Safar1992} 
Furthermore, a nonlocal response perpendicular to the vortices (in-plane) is produced~\cite{Lopez1999,Eltsev2000} as a consequence of the high viscosity of vortices.~\cite{Marchetti1990,Wortis1996,Phillipson1998} Notably, a remote traffic of vortices has been observed as a result of the extremely efficient transfer of local strain induced in one-dimensional vortex arrays that are put in motion elsewhere, at several hundred of vortex spacings.~\cite{Grigorieva2004}  

In this work we focus on the extremely-layered Bi$_2$Sr$_2$CaCu$_2$O$_{8+\delta}$ vortex matter and apply local Hall magnetometry techniques to reveal relevant features of the phenomenology of the first-order transition in vortex matter. 
The detection of the transition line in the low- and high-temperature regions of the phase diagram has been performed applying  several magnetic measurement techniques.~\cite{FOT-Pastoriza,FOT-Zeldov,Ando1999,Avraham,Menghini2003,Beidenkopf2005,Dolz2014,Dolz2015} In the high-temperature region, even a small  $\Delta B$ of a few tenths of a Gauss can be resolved in dc magnetization loop measurements or applying ac  techniques.~\cite{Dolz2014} In the low-temperature region, the first-order transition line bends, significantly decreasing its slope.~\cite{Khaykovich1997,Fuchs1998} In this temperature range the transition manifests as a peak in critical current~\cite{Dewhurst1996,Khaykovich1997} and as a widening of the dc magnetization loops.~\cite{Khaykovich1997,Avraham,Dolz2014} In contrast, the detection of the first-order transition in the intermediate temperature region has been so far elusive by applying standard magnetic measurement approaches.~\cite{Beidenkopf2005,Dolz2014}

Here we present a novel probe of the first-order transformation in layered vortex matter at intermediate temperatures based in direct and remote transmittivity measurements. First, we report a two-step screening of the  ripple field in direct (regular) transmittivity measurements~\cite{Dolz2014} for
the intermediate temperature and field region of the phase diagram. The data present a frequency-dependent first screening step and a frequency-independent second screening step at smaller temperatures.  Second,  we present results probing the remote screening response of the vortex lattice to a local excitation generated away from the detection sensor and pay particular attention to the data in the vicinity of the first-order transition at intermediate temperatures. We generate a ripple field in one location and measure the crossed transmittivity in a remote location of the sample placed  at thousands of lattice spacing distances from the local excitation.  We present evidence of a new nonlocal effect in vortex matter associated to a remote screening of an ac field that echoes the second screening step at roughly the same temperature than in direct screening measurements. Considering results in direct and remote transmittivity data, we provide strong support that the second screening step is a probe of the first-order transition in the intermediate temperature range.

\section*{\label{sec:Exper}EXPERIMENTAL}

\begin{figure}[h]
    \centering
    \includegraphics[width=0.9\columnwidth]{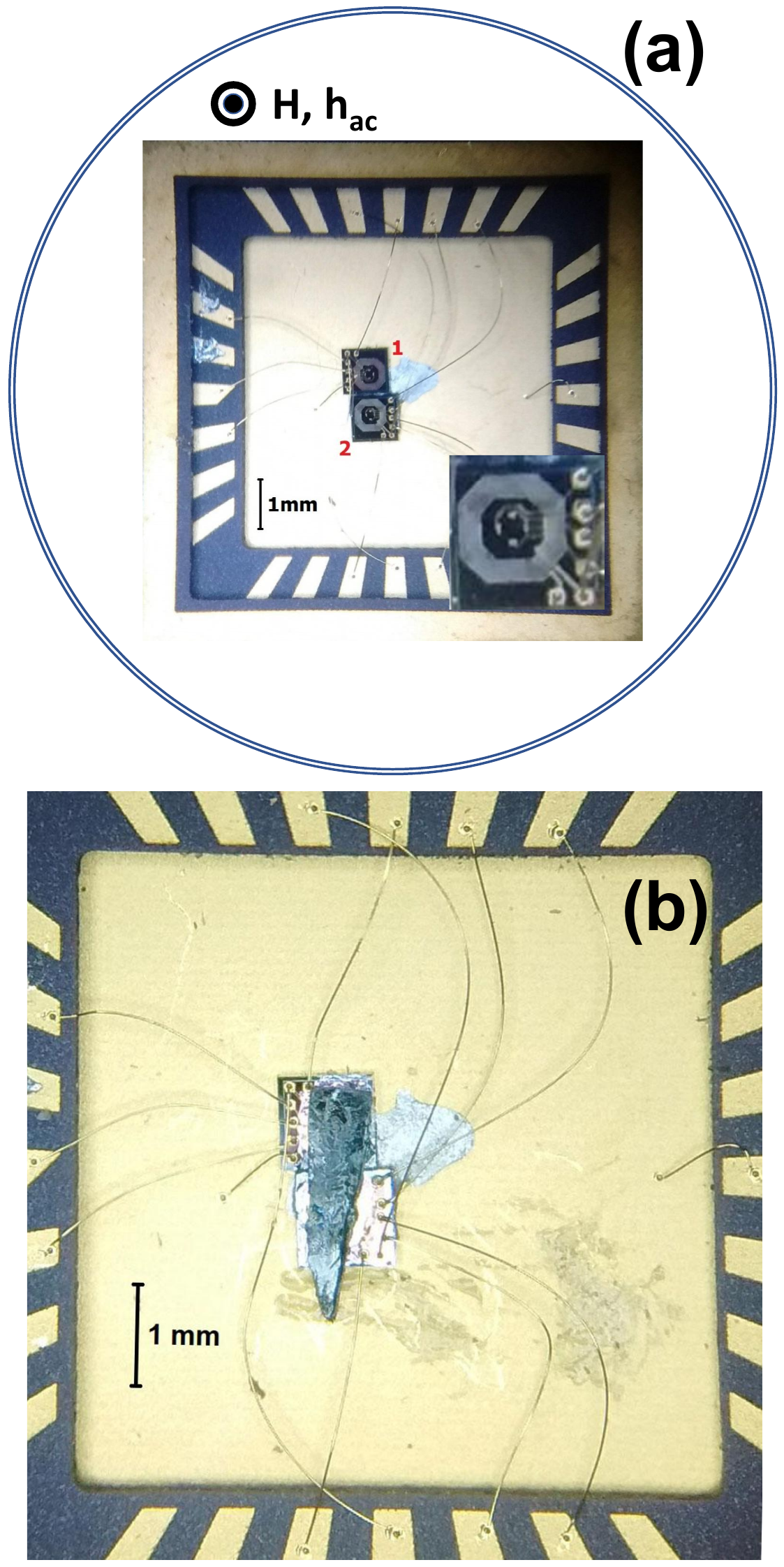} 
    \caption{(a) Pair of Hall sensors and on-chip coils (1 and 2) used for performing Hall magnetometry measurements. The dc field $H$, as well as the $h_{\rm ac}$ excitation, are applied perpendicular to the plane of the chips where lay the sensors. In the \textit{global excitation} mode, $h_{\rm ac}$ is  generated by a  coil external to the chips (see blue lines on the schematics), while in the \textit{local excitation} mode $h_{\rm ac}$ is generated by one of the smaller on-chip coils. Insert: Centered inside each one of the on-chip coils there is a Hall probe with a $9\times9$\,$\mu$m$^2$ detection area, that measures the stray field of the sample. (b) Large Bi$_2$Sr$_2$CaCu$_2$O$_{8+\delta}$ single crystal studied, mounted on top of the two probes with its $c$-axis parallel to the applied fields. }
    \label{fig:FotoMuestra}
\end{figure}

We perform local Hall magnetometry measurements~\cite{ac,Dolz2014,Dolz2015}  on a large nearly optimally-doped Bi$_2$Sr$_2$CaCu$_2$O$_{8+\delta}$ single crystal with $T_{\rm{c}} \simeq$ 85\,K grown by the traveling-solvent floating-zone method.~\cite{Li1994}  The platelet-like crystal is particularly large ($\sim 2$\,mm long and $40$\,$\mu$m thick) as to be mounted on top of a pair of chips  containing each a Hall probe with $9\times9$\,$\mu$m$^2$ detection area surrounded by an on-chip coil. The sensors are fabricated via photolithography onto AlGaAs/InGaAs/GaAs pseudomorphic heterostructures and have a high sensitivity of $\sim 146$\,m$\Omega$/G. The on-chip coils are made up of 7-turn 0.6\,$\mu$m thin Au film deposited over a silicon oxinitride dielectric layer, with a radius ranging from 200 to 320\,$\mu$m and a field intensity in the center of 173 Oe/A. The two chips containing the sensors plus the coils are mounted in a sample holder placed inside of an external coil, see Fig.\,\ref{fig:FotoMuestra}.  The
sample, with its $ab$ plane roughly parallel to the  plane of the sensors, is glued with Apiezon-N grease to
improve thermal contact.

We apply dc and ac fields,  $H$ and $h_{\rm ac}$ respectively, both perpendicular to the plane of the sensors (parallel to the $c$-axis of the sample). When applying an ac ripple field $h_{\rm ac}$ as a perturbation of a larger dc field $H$, a superconductor screens the penetration of the ac ripple field. The  Hall probes measure the local stray magnetic induction of the sample in the $c$-axis direction, $B^{\perp}$. We perform dc magnetometry measurements by measuring the sample magnetization, $4\pi M = B^{\perp} - H$.  We also apply two ac measurement modes: The \textit{local excitation} mode with $h_{\rm ac}$ generated by the on-chip coils and the \textit{global excitation} mode with $h_{\rm ac}$ produced by a centimeter-sized coil generating a ripple field in the whole area of the measurement chip, see Fig.\,\ref{fig:FotoMuestra}.  In the ac measurements we simultaneously acquire the first and third harmonics of the
Hall voltage (proportional to the magnetic induction $B^{\perp}$) by means of a lock-in technique.~\cite{Dolz2015} The transmittivity of the vortex system is obtained by normalising the in-phase
component of the first-harmonic signal, $B'$, such that  $T' = [B'(T) - B' (T \ll T_{\rm c}]/[B'(T > T_{\rm c}) - B' (T \ll T_{\rm c})]$. $T'$ is  extremely sensitive to changes in local induction, in particular discontinuities associated with first-order transitions.~\cite{Dolz2014} The magnitude of the third-harmonic signal, $B_{\rm h3}$, is zero when the magnetic response is linear.~\cite{Bean2,Dolz2014} The normalized  magnitude $|T_{\rm h3}|= B_{\rm h3}/[B'(T > T_{\rm c}) - B' (T \ll T_{\rm c}]$ becomes non-negligible when non-linearities in the magnetic response develop.~\cite{Dolz2014} For all our measurements  the phase of the lock-in signal is set to zero at 90\,K  before starting the measurements.

The sample is mounted on top of two chips, each of them containing a Hall sensor surrounded by an on-chip coil.  We present 
transmittivity data following three different measurement protocols. Two of them are referred to as \textit{direct transmittivity} $T'$ measurements since the Hall sensor is detecting the signal within the spatial region where the ac ripple field is generated. We distinguish between direct transmittivity measurements in global and local excitation modes when the $h_{\rm ac}$ is generated by the  external coil or the on-chip coils, respectively. The third protocol is the \textit{crossed transmmittivity} measurement that consists in locally generating $h_{\rm ac}$ in one on-chip coil while measuring the magnetic properties on the sensor located on the other chip, outside of the powered coil. This is not strictly speaking a transmittivity measurement since  the excitation is generated outside of the detection area. We call this magnitude $T'_{\rm cross}$, slightly  abusing the notation, but making clear this signal is also the in-phase component normalized by its high- and low-temperature limiting values.  We follow these three measurement protocols in order to characterize the  screening response of the vortex structure to local and non-local ac excitations. This allows us to provide useful information for the interpretation of the typical features that appear in the ac screening  of vortex matter and their connection to phase transitions.

\section*{RESULTS}

\subsection{First-order vortex phase transition probed by dc and global ac  Hall magnetometry}

\begin{figure}[ttt]
    \centering
    \includegraphics[width=0.7\columnwidth]{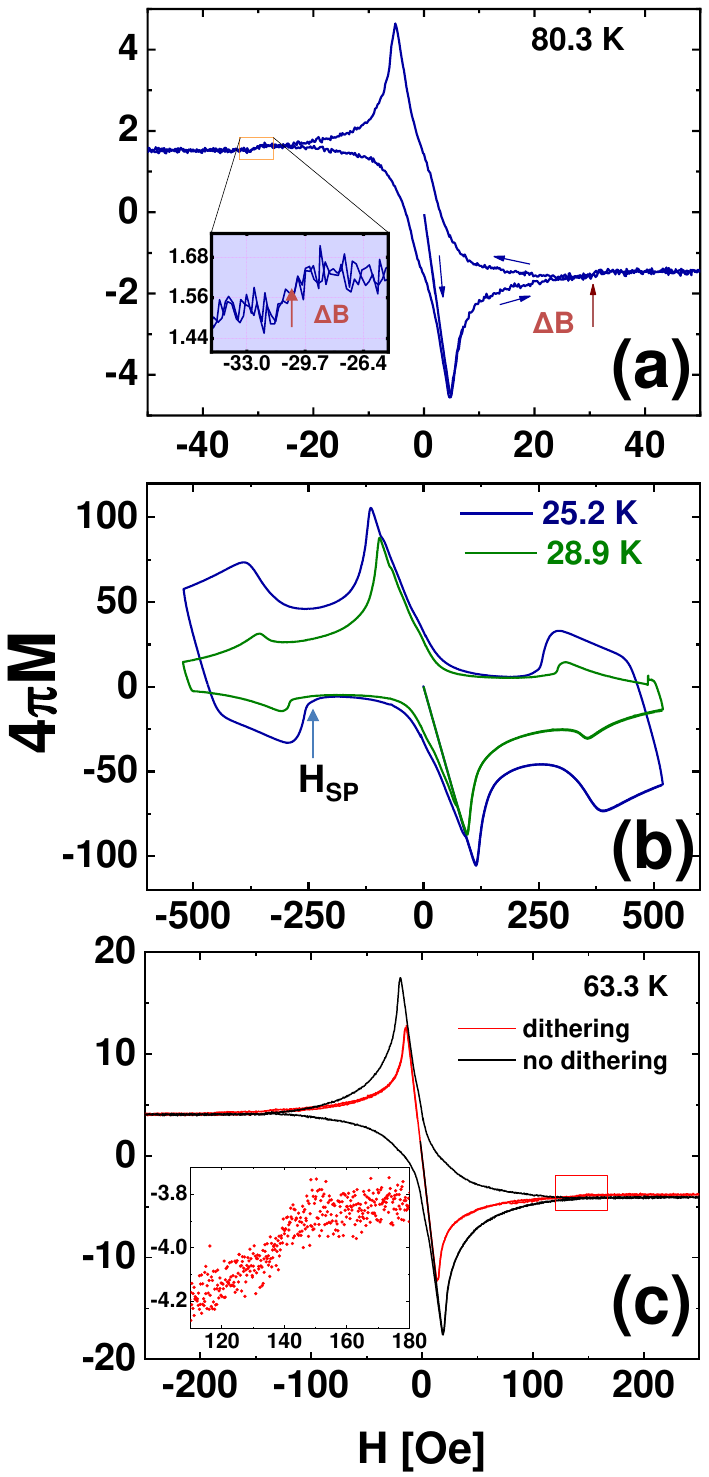}
       \caption{Typical dc magnetization loops of vortex matter in Bi$_2$Sr$_2$CaCu$_2$O$_{8+\delta}$ measured by local Hall magnetometry at different temperature ranges. (a) In the high-temperature range, $T>60$\,K, the reversible magnetization presents a sub-Gauss jump in induction, $\Delta B$, a manifestation of the first-order vortex melting transition. This jump is zoomed in the insert for the negative $H$ branches. (b) In the low temperature range, $T<30$\,K, irreversibility is important and the magnetization loops widen. The order-disorder transition manifests as a local increase in the widening of the loop indicated as $H_{\rm SP}$. (c) Loop at intermediate temperatures for the regular and dithering measurement protocols (in-plane alternating pulsed field of 30\,Oe during 3\,msec, 0 during 7\,msec and -30\,Oe during 3\,msec, repeated cyclically during the whole measurement).  In the latter case hysteresis is suppressed and the magnetization jump associated with the first-order  transition is made evident also in the intermediate temperature range.}
    \label{fig:dc}
\end{figure}

We start by revisiting how the first-order phase transition is detected in vortex matter applying dc magnetization techniques. In this type of measurements, a sweep of the applied field $H$ results in hysteresis loops $M(H)$ such as the ones presented in the three panels of Fig.\ref{fig:dc}.
Panel (a) shows typical results within the high-temperature range: An irreversible response is detected at low fields but the loop closes when increasing field indicating the response becomes reversible. Within this reversible region, a small jump in $M$ coming from a sudden change  in vortex density, $\Delta B$, is detected  when varying $H$, see for instance the data at 80.3\,K in the insert to Fig.\,\ref{fig:dc}(a). This modest jump of at best a tenths of Gauss is the fingerprint of the first-order transition entailing a change in vortex density between the high-temperature liquid phase and the low-temperature solid vortex phase.~\cite{Avraham,Dolz2014}

 For the low-temperature range smaller than $\sim 30$\,K irreversibility in the magnetic response becomes more important and the magnetization loops widen up to two orders of magnitude in comparison to the high-temperature data, see Fig.\,\ref{fig:dc}. The enhancement of irreversibility at low temperatures is due to the growing relevance of bulk pinning on cooling. In this temperature range the first-order transition is detected as a sudden change in the slope of $M$ that further enhances the width of the loop, see feature $H_{\rm SP}$ indicated in Fig.\,\ref{fig:dc} (b). The local minimum (maximum) of the ascending (descending) branches of magnetization becomes more pronounced upon cooling. This phenomenology is known as the second peak effect and entails a peaked-like enhancement of the sustainable current  on increasing field or temperature.~\cite{CreepRegimes,2peak-Ord-Dis}   We consider the field of the onset of the second peak effect as the  first-order transition field, see the arrow in Fig.\,\ref{fig:dc} (b). In order to draw the $H_{\rm SP}$ line in the vortex phase diagram we average the field location of the onset in the four measured branches and consider the statistical dispersion as its uncertainty. 

\begin{figure}[ttt]
    \centering
    \includegraphics[width=\columnwidth]{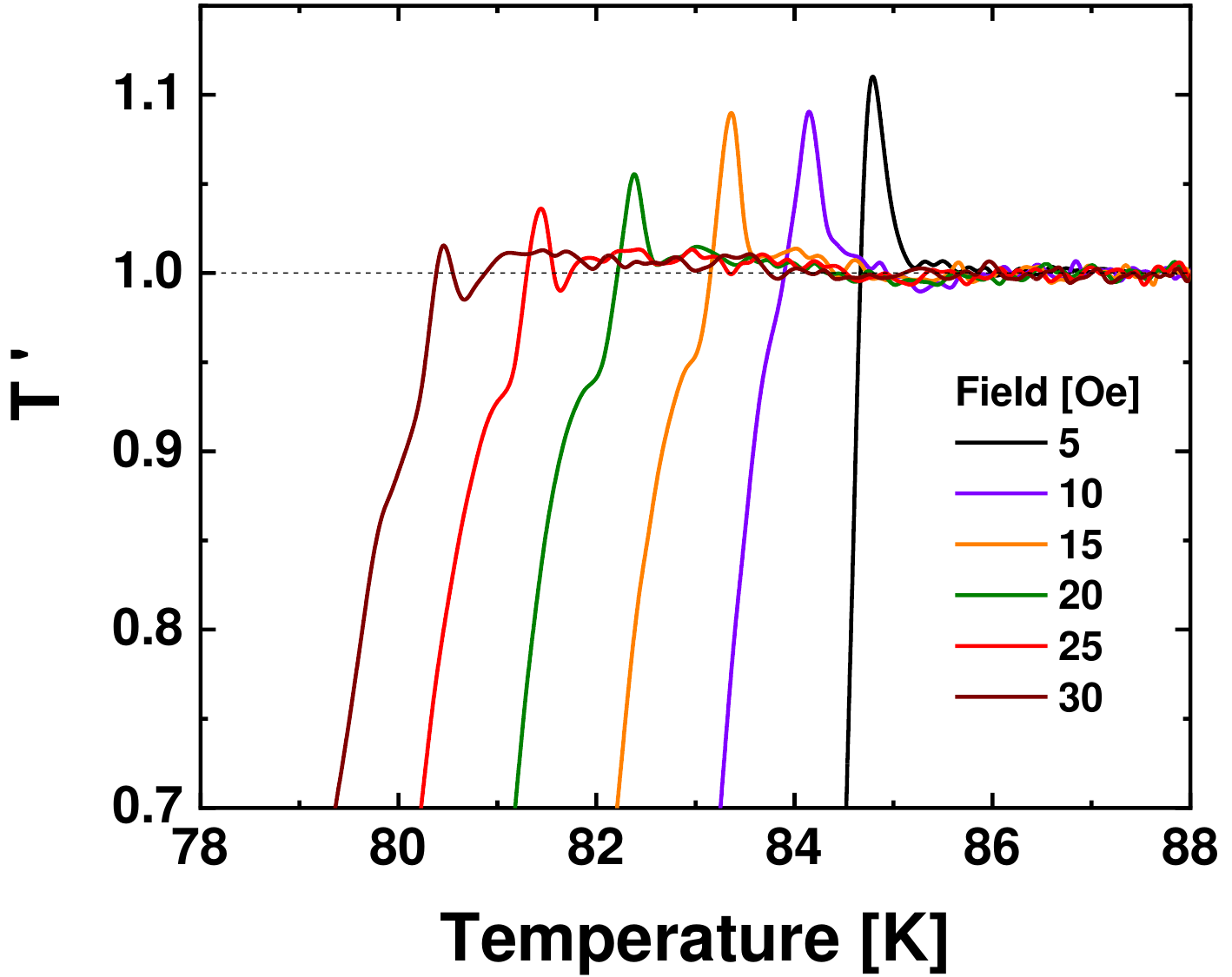}
    \caption{ac transmittivity $T'$ as a function of temperature in the low-field range. The so-called paramagnetic peak entailing values of $T'>1$ is a signature of the jump in $B$ produced at the   first-order melting transition. The measurements were performed at the dc fields indicated in the legend, using an ac ripple field of  0.4\,Oe$_{\rm rms}$ and 17.1\,Hz.}
    \label{fig:ac}
\end{figure}

The $H_{\rm SP}$ feature in dc magnetization loops is harder to detect at intermediate temperatures between $\sim 30$\,K-$65$\,K,~\cite{Dolz2014} where the $\Delta B$ is rather small as to be distinguished from the experimental resolution. This limitation has been overcome by applying a non-standard magnetization measurement performed while applying  a kHz-pulsed alternating dithering field in the in-plane direction concomitantly with the dc $H$ field applied in the c-axis.~\cite{Avraham} 
By applying this protocol it is possible to equilibrate  vortex matter and thus reduce irreversibility effects coming from pinning in the samples.~\cite{Avraham} Irreversibility is diminished in the dithering magnetization loops in comparison with regular dc measurements, see for instance data at 63.3\,K in panel (c) of Fig.\ref{fig:dc}. Thus, this method enables the detection of a jump $\Delta B_{\rm dith}$ in the reversible region of the dithering loop at the intermediate temperature range, see insert to Fig.\ref{fig:dc} (c). As useful as the results obtained with this method are, adopting this protocol requires applying large and well oriented in-plane ac fields on top of the c-axis-oriented dc field.

At small fields, the in-phase response to an applied ripple field shows a paramagnetic peak recorded either as a function of temperature or magnetic field, due to the jump in $B$ at the first-order transition.~\cite{ac,FOT-Pastoriza,FOT-Zeldov}   For small ripple fields, the height of the peak at the temperature $T_{\rm FOT}$ is approximately $T'(T_{\rm FOT})  = 1 +  2\Delta B / \pi h_{\rm ac}$, thus proportional to the entropy-jump at the transition.~\cite{Dolz2014}
 Figure\,\ref{fig:ac} shows a set of measurements of $T'$ as a function of temperature for various fields in the low-$H$ range. All curves present a transmittivity value close to one at high temperatures, a paramagnetic peak where $T'>1$,  and a suppression of transmittivity at lower temperatures associated with the screening of the ac field.  The temperature location of the peak does not depend on the frequency nor the amplitude of $h_{\rm ac}$, as expected from a first-order transition.~\cite{Dolz2014} On increasing $H$ the peak appears at systematically lower temperatures and eventually fades into the screening region of the curve making it difficult to ascertain its location. For the sample studied here we are able to detect its location for $H \lesssim 50$\,Oe. For larger fields, the decreasing amplitude of the paramagnetic peak and its shift to lower temperatures where screening produces a sudden step in $T'$ prevents its detection above the experimental noise of the ac measurement. Thus, this global excitation ac technique looses resolution to track the first-order transition line from a paramagnetic peak in the intermediate temperature region.

\begin{figure}
    \centering
    \includegraphics[width=1\columnwidth]{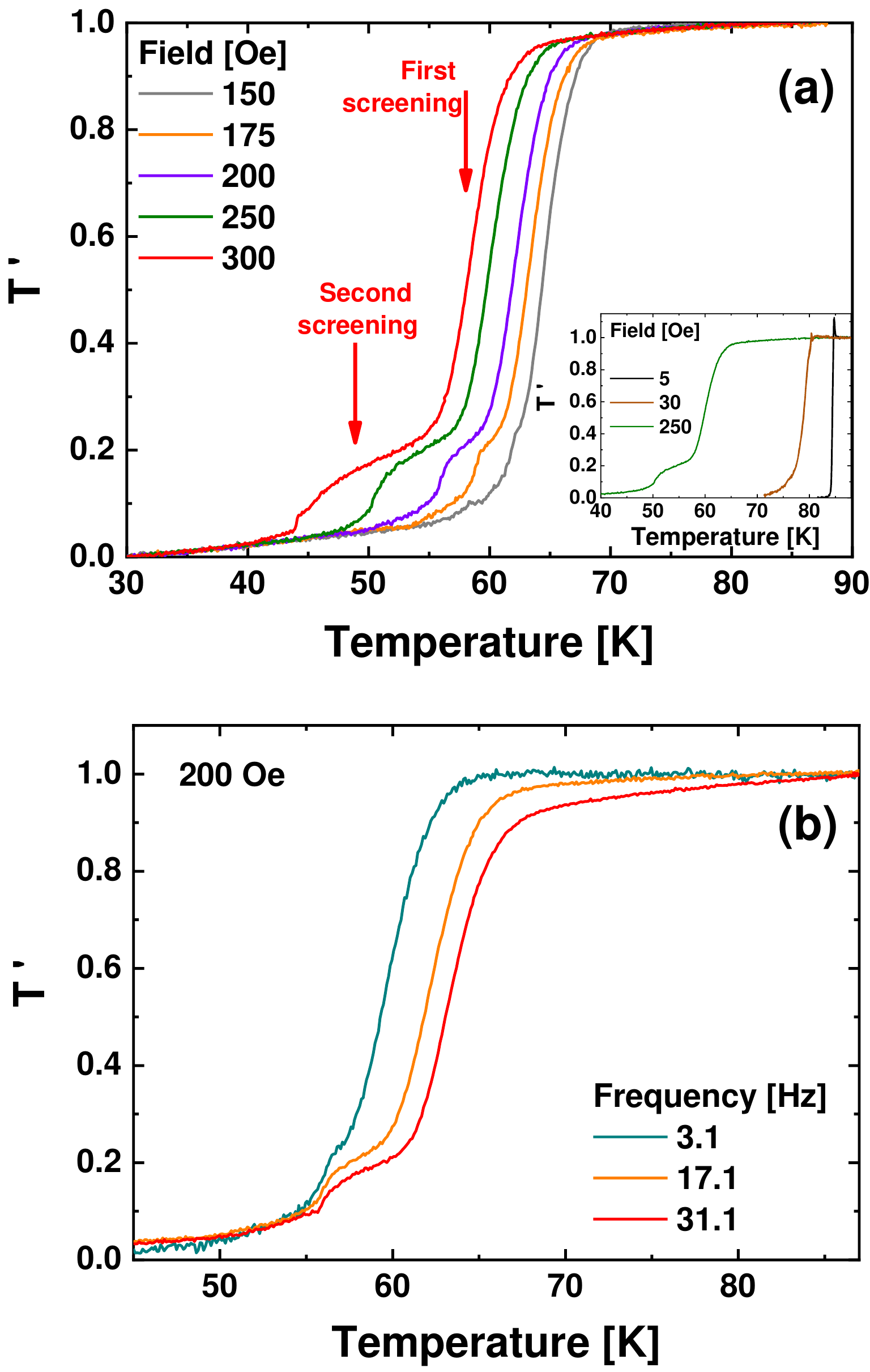}
    \caption{Transmittivity measurements in Bi$_2$Sr$_2$CaCu$_2$O$_{8+\delta}$ on decreasing temperature following a field-cooling process at a field $H$. (a) Main: data for $H$ in the high-field range from 150 to 300\,Oe showing a first and a second screening response. Insert: Comparison between data obtained at high and low fields. All data in this panel come from measurements performed at a ripple field of 17.1\,Hz and 0.4\,Oe$_{\rm rms}$.  (b) Transmittivity when field-cooling at 200\,Oe applying a ripple field with various frequencies and 0.4\,Oe$_{\rm rms}$.   The temperature location of the first screening shifts with frequency  while that of the second screening remains unchanged. }
    \label{fig:2Feat}
\end{figure}

 However, on further increasing field, for $H > 150$\,Oe the  ac signal displays a peculiar behavior of the screening  stage that ubiquitously persists in the intermediate field range up to $H_{\rm SP}$. The transmittivity data of Fig.\,\ref{fig:2Feat} shows that, after following a first screening at high temperatures, the signal presents a \textit{plateau} and then a second screening step at lower temperatures. Panel (a) of this figure shows $T'$ as a function of temperature for various fields ranging from 150 \,Oe to 300\,Oe, performed with an ac excitation of amplitude 0.4\,Oe rms and a frequency of 17.1\,Hz.  On increasing field, the second screening step is sharper and the temperature range occupied by the plateau becomes wider.  This  contrasts with the phenomenology at fields $H<50$\,Oe, for which the screening occurs in a single step, see insert to Fig.\,\ref{fig:2Feat} (a).

\begin{figure}
    \centering
    \includegraphics[width=\columnwidth]{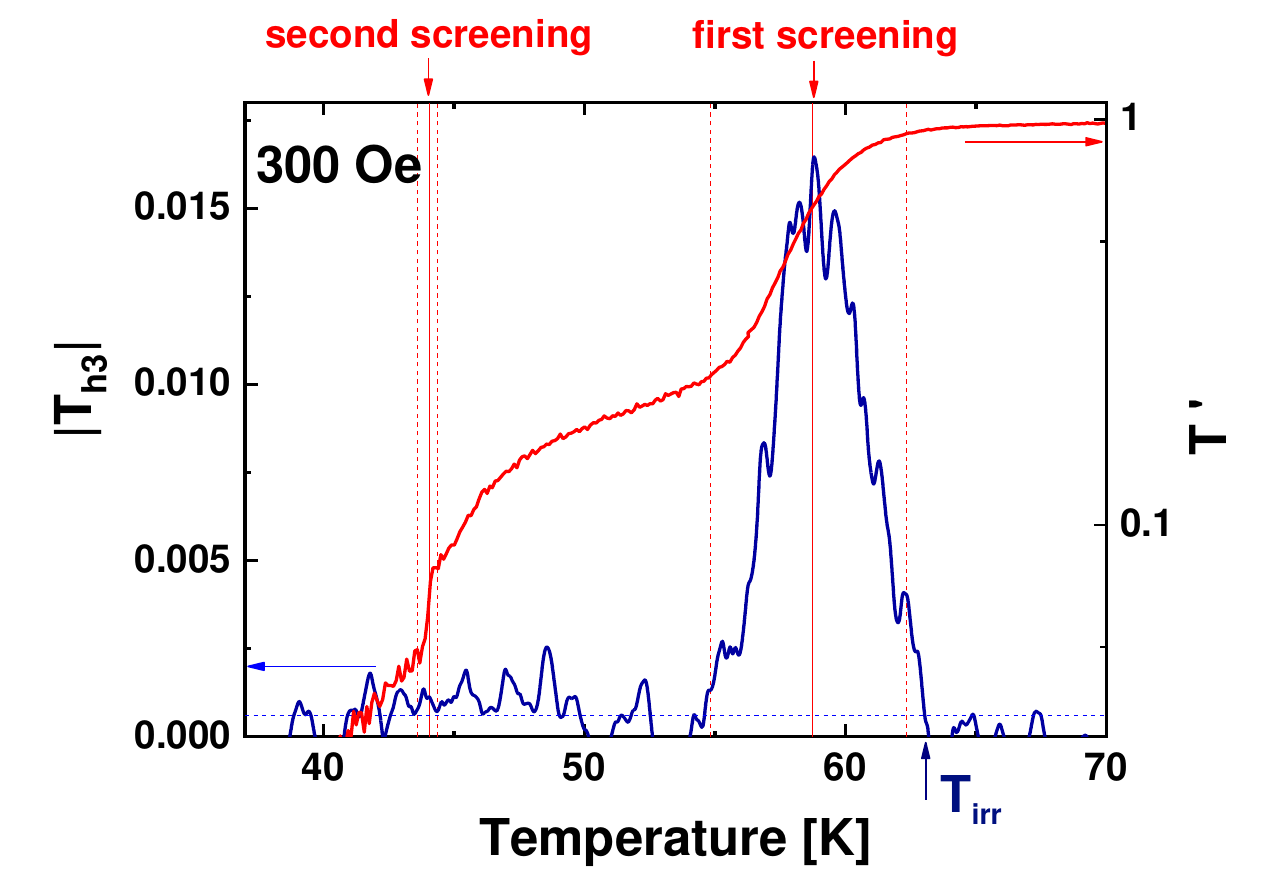}
    \caption{Normalized third harmonic signal $|T_{\rm{h3}}|$ (left axis) and transmittivity (right log-scale axis) as a function of temperature, recorded simultaneously while following a field-cooling protocol. Data for an applied dc field of 300\,Oe and a ripple field of  17.1\,Hz and 0.4\,Oe rms.
   The midpoints of both screening steps are indicated with red arrows. The midpoint in the first screening in transmittivity coincides with the location of the larger peak in  $|T_{\rm{h3}}|$ while the midpoint in the second screening is within a faint and broad peak registered at smaller temperatures. A horizontal dashed blue line indicates the upper noise level in the third harmonic signal. The irreversibility temperature $T_{\rm irr}$ at which the magnetic signal becomes non-linear is estimated from the onset of 
    $|T_{\rm{h3}}|$ above the upper noise level on cooling, see blue arrow. }
    \label{fig:Th3}
\end{figure}

Fig.\,\ref{fig:2Feat} (b) shows typical transmittivity data at  $H=200$\,Oe, an example of the typical phenomenology observed for the first and second screening steps detected at intermediate $H$ fields when changing the frequency of the $h_{\rm ac}$ field. The first screening observed at high temperatures has a strong dependence with the amplitude and frequency of the excitation. For example, the first screening shifts to higher temperatures on increasing frequency or decreasing the amplitude of the excitation. We estimate the typical temperature of the first screening step from the middle point between the onset and the end of the first sudden change of transmittivity on cooling, see dotted and full red lines in Fig.\,\ref{fig:Th3}. In contrast, the characteristic temperature of the second screening step is robust against changes in both the frequency and amplitude of the excitation, see Fig.\,\ref{fig:2Feat} (b). The magnitude of the signal of the second screening changes with frequency since the plateau has a smaller height on increasing frequency, but the sudden second step in $T'$ on cooling is observed at similar temperatures irrespective of the frequency of the excitation. Following a similar criteria for estimating the typical temperature of the second screening, see dotted and full red lines at the left of Fig.\,\ref{fig:Th3}, we see no changes with frequency in the typical temperature of the second screening within the experimental uncertainty.

\begin{figure}
    \centering
\includegraphics[width=\columnwidth]{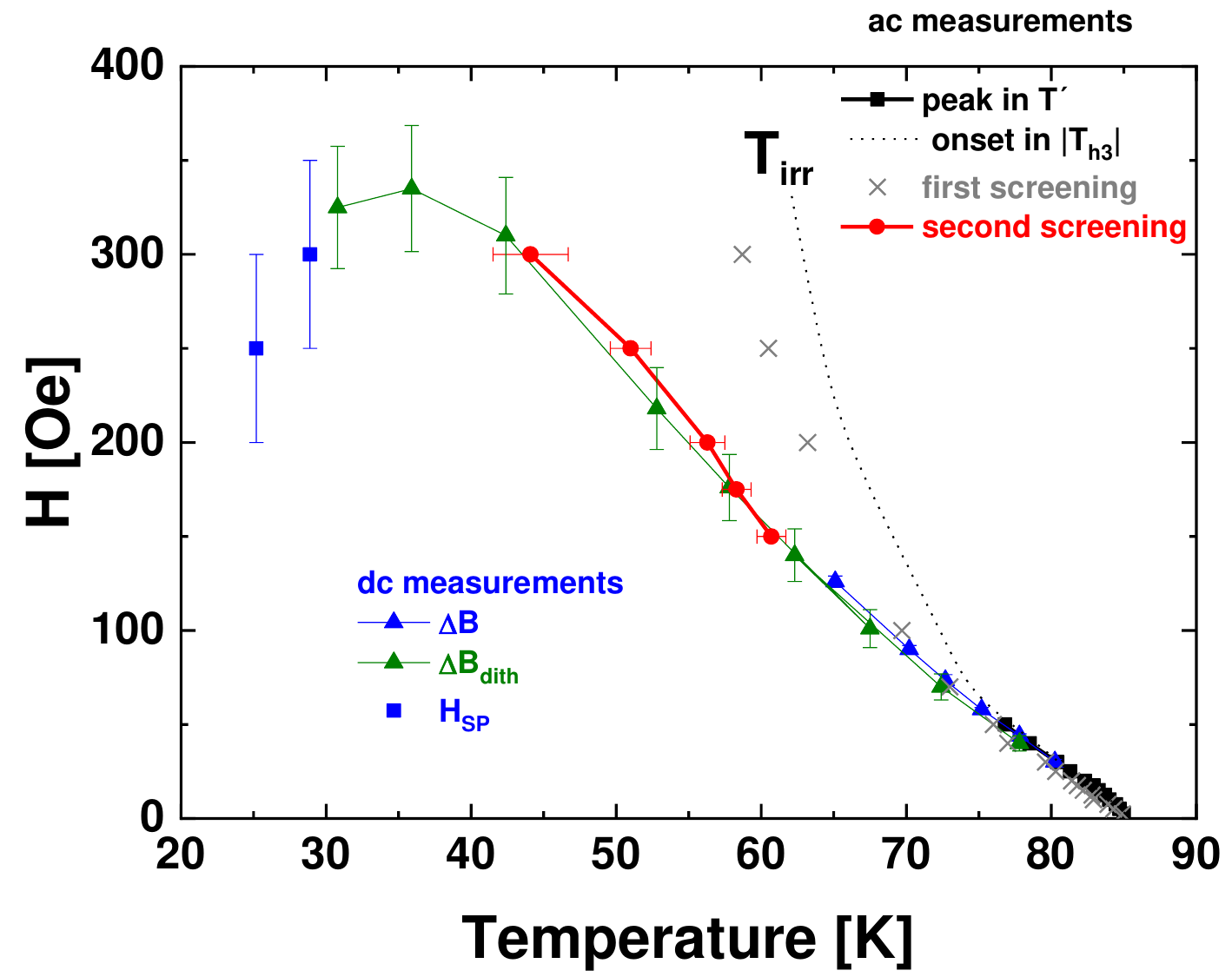}
    \caption{Phase diagram of vortex matter in Bi$_2$Sr$_2$CaCu$_2$O$_{8+\delta}$ obtained from  different ac and dc magnetic measurements. The first-order melting is tracked in the low-field or high-temperature region by following the paramagnetic peak in $T'$ (black squares), and the $\Delta B$ jumps measured in dc magnetization  (blue triangles) and  dithering (olive triangles) loops. The continuation of this line at lower temperatures is the order-disorder first-order transition $H_{\rm SP}$ detected via the second peak effect in dc magnetization loops (blue squares). The onset of irreversible magnetic response on cooling is detected from the onset (dotted black line) of the $|T_{\rm h3}|$ signal  that presents a maximum at around half of the first-screening step (gray crosses). These features are obtained from measurements at 17.1\,Hz but their temperature location strongly depends on the frequency of the excitation $h_{\rm ac}$. In red circles we indicate the location of the second screening feature detected in $T'$ versus temperature measurements.}
    \label{fig:PhDiag}
\end{figure}

This sudden jump in $T'$ registered at the second screening feature and its independence with the excitation signal suggests it could be associated with in the $B$ \textit{versus} $T$ dependence that could be connected to the first-order transition. In addition,  plotting the second screening typical temperature in a phase diagram together with the temperature-location of $\Delta B$ from dc magnetization data (regular and dithering protocols) and the temperature-location of the paramagnetic peak in $T'$  reveals that  the second screening coincides with the first-order line within the experimental error. The error bars for the second screening temperature in the phase diagram come from the half width of the screening step as indicated with dotted lines in Fig.\,\ref{fig:Th3}. Interestingly, the second screening is detected in the intermediate field range where  the sophisticated dithering protocol is able to reveal the first-order transition, connecting the region of detection of the paramagnetic peak at low fields and the second peak effect at high fields. This second screening step in transmittivity \textit{versus} temperature is detected in the same $T$ and $H$ range as the frequency-independent $H_{\rm step}$ feature observed in $T'(H)$ loops and reported in Ref.\,\onlinecite{Dolz2014}. Therefore, the second screening step in regular ac protocols with $H$ applied along the c-axis (either detected varying temperature or field) probes the first-order transition in a field and temperature region where other detection protocols are insensitive, with exception of the sophisticated dithering method.

The screening of the ripple field is detected not only by measuring the transmittivity, related to the first harmonic of the magnetic response of vortex matter to an ac field,  but also by measuring the normalized third harmonic of that signal, $|T_{\rm h3}|$. A non-negligible value of $|T_{\rm h3}|$ has origin in a noticeable non-linear response of vortex matter generally associated with an irreversible magnetic response. Figure\,\ref{fig:Th3} shows an example of a typical $|T_{\rm h3}|$ \textit{vs.} temperature curve on field-cooling at 300\,Oe (blue data). On starting cooling, the third harmonic signal becomes non-negligible above the noise level (blue dashed line) below a temperature $T_{\rm irr}$ (see arrow). This is the irreversibility line that marks the onset of irreversible magnetic response due to bulk pinning and surface barriers, see dotted black line in the phase diagram of 
Fig.\,\ref{fig:PhDiag}. This line has a strong frequency dependence and signals a crossover from a reversible liquid at high temperatures to an irreversible liquid at low temperatures.~\cite{Dolz2014} On  cooling, $|T_{\rm h3}|$ develops a peak with its maximum located at roughly the same temperature as the characteristic first screening temperature, see Fig.\,\ref{fig:Th3}.  On further cooling, the typical temperature of the second screening step does not replicate in a particular structure in the $|T_{\rm h3}|$ signal. Therefore, the third harmonic signal presents  signatures associated only with the first screening process of the ac field.  The temperature location of the peak in $|T_{\rm h3}|$, coinciding within the error with the typical temperature of the first screening observed in $T'$, is indicated in the phase diagram with gray crosses.

We stress that, while the first screening temperature depends on frequency,
the second screening temperature is not altered by changes of frequency nor amplitude of the ac excitation and overlaps the jump in $B$ detected in dithering experiments associated with the first-order transition. Therefore, the second screening process can be used as a probe for tracking the first-order phase transition in the intermediate temperature and field region where the regular ac and dc magnetic techniques with $H$ and $h_{\rm ac}$ applied collinear loose resolution to detect the $\Delta B$ associated with this transition.

\subsection{Remote detection of the two-step ac screening}\label{sec:Crossed_trans}

\begin{figure}
    \centering
    \includegraphics[width=\columnwidth]{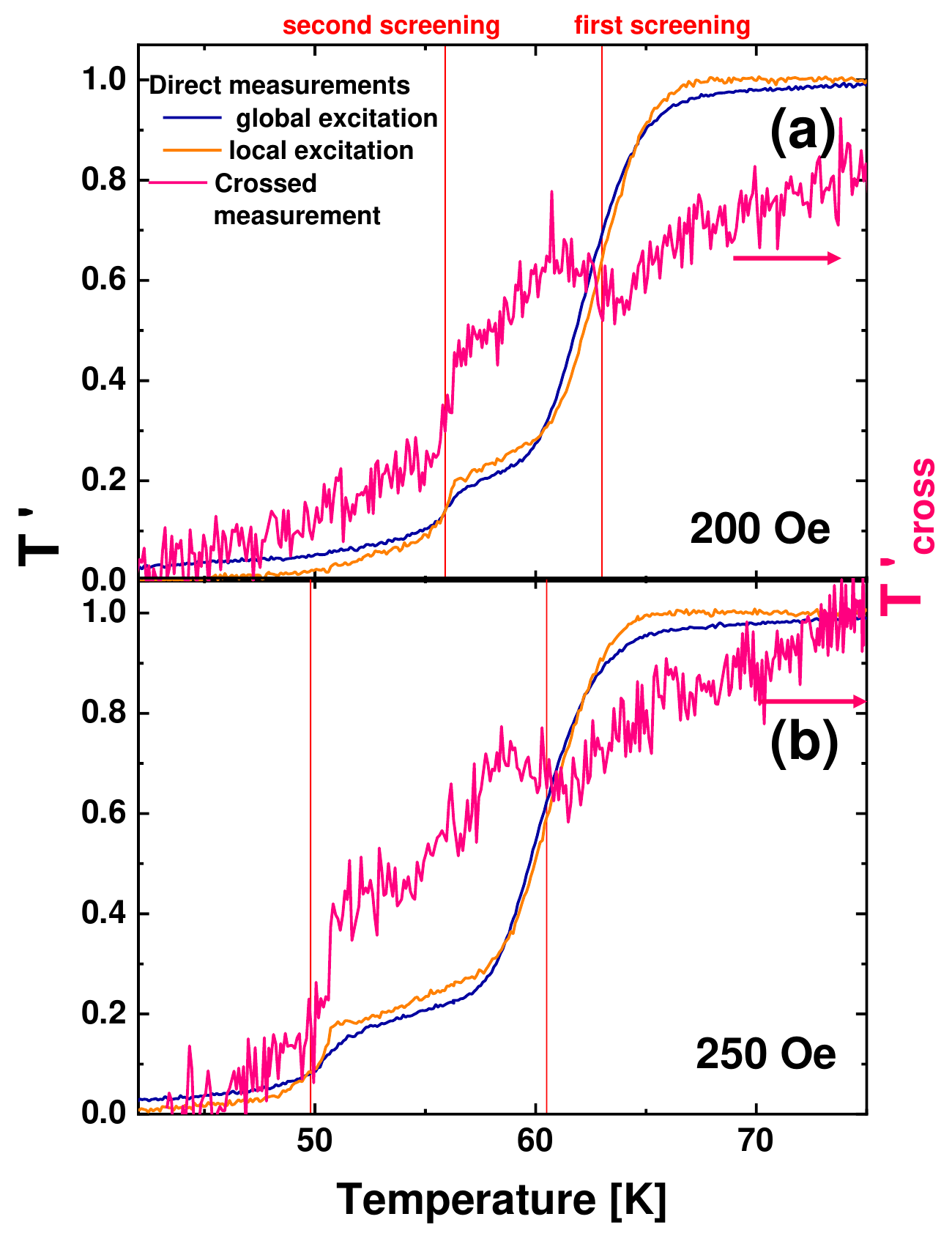}
    \caption{ Transmittivity as a function of temperature while cooling with (a) $H=200$\,Oe and (b) $H=250$\,Oe, with an ac ripple field of 17.7\,Hz and 0.4\,G rms. The results are color coded according to the three measurement protocols: direct measurements with global and local excitations and crossed measurements. Vertical red lines indicate the detection of the first and second screening in transmittivity. Data measured with an ac ripple field  of  17.1\,Hz and 0.4\,G rms.}
    \label{fig:Trans-2f}
\end{figure}

We devise a crossed transmittivity measurement protocol to probe the remote screening response to a local ac excitation generated thousands of vortex lattice spacings away from the detection sensor. This protocol has been applied in order to 
identify the possible mechanisms at the origin of the second screening step that we propose to use in order to track the first-order transition. Figure\,\ref{fig:Trans-2f} shows examples of direct $T'$ and crossed $T'_{\rm cross}$ transmittivity data for 200\,Oe and 250\,Oe applied fields. Data for direct measurements with $h_{\rm ac}$ generated globally (blue lines) with the external coil  or locally (orange lines) with the on-chip coil surrounding the same detection Hall sensor are similar for all the studied fields. In particular, the characteristic temperatures of the first and second screening steps remain unaltered by generating the excitation locally or globally. We are able to detect a signal $T'_{\rm cross}$ (pink lines) from crossed measurements   performed on the Hall sensor located outside and away ($\sim 700$\,$\mu$m) from the on-chip coil where $h_{\rm ac}$ is locally generated. The absolute value of the signal (before normalizing to obtain $T'_{\rm cross}$) is one order of magnitude smaller than in the case of direct measurements, and thus, these magnitude is noisier than $T'$.

The behavior of $T'$ and $T'_{\rm cross}$ with temperature are rather different.  
At the  typical temperature of the first screening $T'_{\rm cross}$ registers a smooth decrease on cooling instead of a jump or step-like behavior as in direct transmittivity measurements. This  is actually originated by a reinforcement of the signal within the \textit{plateau} region between the first and second screening temperatures in direct transmittivity measurements, see red lines in Fig.\,\ref{fig:Trans-2f}).  But interestingly, at the second screening temperature the crossed transmittivity presents a step-like behavior as in the case of direct transmittivity, reinforcing the idea that this feature is a probe of the first-order transition.

\section*{DISCUSSION}

As mentioned, Fig.\,\ref{fig:PhDiag} presents  a phase diagram constructed from the different ac and dc measurement protocols used in this work, namely regular ac, crossed ac, regular dc and dithering dc. The temperature location of the particular features associated with the $\Delta B$ detected in the dc and ac measurements reported in the literature or proposed in this work defines the first-order transition line in vortex matter. The location of these features is frequency- and amplitude-independent in ac measurements. In the current work  we show that the second screening step is a sensitive method to detect the first-order transition by means of regular ac magnetometry  in the intermediate temperature region. The results obtained from crossed transmittivity measurements reinforce this last assertion: A jump in $T'_{\rm cross}$ is detected at a frequency- and amplitude-independent temperature that coincides with the  characteristic temperature of the second screening step detected in $T'$.

 Surface barriers, of either Bean-Livingston~\cite{Bean-Livingston} or geometrical~\cite{Zeldov1994} type, hinder the entrance and exit of vortices from the sample. However, they do not prevent the internal rearrangement of the vortices already located inside of the sample. The onset of screening due to surface barriers, relevant at high temperatures, produces a decrease in direct transmittivity data (both with local or global excitations) since the penetration of new vortices is hindered. The modulus of the crossed signal is expected to increase, anti-correlated  with the direct transmittivity data, due to local rearrangements of  vortices as a result of generating the excitation in a region of the sample away from the detection area. This is indeed observed in the data of Fig.\,\ref{fig:Trans-2f} in the temperature region between the first and second screening steps. When bulk pinning is dominant, it affects equally the direct and remote screening signals. As a consequence,  a  decline of both, direct and crossed transmittivity, is expected when traversing the first-order transition on cooling. This is indeed observed below the second screening temperature where the three measurement protocols show a decrease in transmittivity towards lower temperatures, see Fig.\,\ref{fig:Trans-2f}.

Therefore, the two main sources of screening affect the direct and crossed transmittivities in different ways. Taking this into account allows us to explain the reinforcement in crossed transmittivity data observed in the temperature range between the first and second screening temperatures. At temperatures higher than the first screening step, surface barriers are dominant and vortices are hindered to enter or exit the sample in order to screen the ac excitation. Then, on cooling through 
the fist screening temperature, the non-local excitation in crossed transmittivity experiments can only be screened by local rearrangements of vortices inside of the sample. In this local rearrangement vortices migrating towards (from) the region of the sample located below the excitation coil are coming from (going to) other parts of the sample and thus the density in the detecting sensor is smaller (larger). As a consequence, in the temperature range between the first screening and the second screening step the $T'_{\rm cross}$ signal is larger and in counterphase than the direct $T'$ signal. Once the second screening step temperature is reached on cooling, if this step is associated to the first
order transition as we argue, then this feature is due to a global change in vortex density in the whole sample: It  entails the entrance and exit of vortices, resulting in a decrease of both, $T'$ and  $T'_{\rm cross}$.
These two trends are indeed observed in the data of Fig.\,\ref{fig:Trans-2f}. Thus the temperature dependence of the signal in crossed transmittivity experiments provides further evidence that the second screening step is associated to the first order transition in layered vortex matter.

For a large region of the phase diagram, the first-order transition occurs at temperatures below the first screening. Thus, a region of the liquid vortex phase presents a non-linear behaviour, an issue already discussed in Ref.\,\onlinecite{Dolz2014}. According to the interpretation of the screening mechanisms put forward here in view of the crossed transmittivity data, this non-linear response is quite likely ascribed to the effect of surface barriers contributing to the screening of the ripple field.

\section*{CONCLUSION}

We detect a two-step screening of an ac excitation field for Bi$_2$Sr$_2$CaCu$_2$O$_{8+\delta}$ vortex matter when the dc applied field is over 150\,Oe. In view of this result, we devised a crossed  transmitivitty experiment to probe the remote response of the vortex lattice to a local excitation generated away from the detection region. The temperature location of the frequency- and amplitude-independent second screening feature coincides within the error with the location of the first-order transition line that we detect by means of  non-standard magnetic dithering techniques.~\cite{Avraham,Beidenkopf2005}  
The remote screening response echoes
the second screening step at roughly the same temperature as the one determined by exciting and detecting at the same location. This phenomenology is observed in the intermediate temperatures and fields range. Therefore, the second screening feature in regular and crossed transmittivity experiments is a sensitive probe to detect the first-order transition in high-temperature layered vortex matter in a phase diagram region where its detection was previously elusive. Furthermore, our remote magnetometry data at the vicinity of the first-order transition temperature  suggests that the elasticity of the solid vortex structure with a finite shear modulus is at the origin of an efficient method to propagate the rigidity of the less symmetric phase and remotely inducing a jump in $B$ outside of the excitation area.

\section*{Acknowledgments}

We thank Vincent Mosser for the VNEMETIC Hall sensors with integrated coils whose characteristics allowed the measurements 
reported in this paper.

\bibliography{biblio}

\end{document}